# A 100% Renewable Energy System: Enabling Zero CO2 Emission Offshore Platforms

Cunzhi Zhao
*Student Member, IEEE*
Department of Electrical and Computer Engineering
University of Houston
Houston, TX, USA
czhao20@uh.edu

Xingpeng Li
*Senior Member, IEEE*
Department of Electrical and Computer Engineering
University of Houston
Houston, TX, USA
xli82@uh.edu

*Abstract*— The total electricity consumption from offshore oil/gas platforms is around 16 TWh worldwide in 2019. The majority offshore platforms are powered by the diesel generators while the rest mainly uses gas turbines, which emits large amounts of CO2 per year. The fast development of offshore wind turbines (WT) can potentially replace traditional fossil fuel based resources to power offshore loads. Thus, a novel offshore hybrid renewable energy sources (OHRES) system is proposed to enable a zero CO2 emission offshore platform mitigating climate change. Battery energy storage system (BESS) and hydrogen energy storage system (HESS) are considered to mitigate the fluctuation of wind power in the proposed OHRES system. Resilience models are designed to enhance the resilience of the proposed OHRES system with extra energy stored in BESS and/or HESS. Case studies demonstrate the feasibility of the proposed OHRES system to power offshore platforms. The economic analysis reports the planning cost for the proposed OHRES system under different resilience levels, which may benefit the decision to balance the carbon emission and investment cost.

## Nomenclature

*Sets*
$T$ — Set of time intervals.

*Indices*
$t$ — Time interval t.

*Parameters*
$c_{WT}^{Capital}$ — Capital cost for each wind turbine unit.
$c_{BESS}^{Capital}$ — Capital cost for BESS per MWh.
$c_{El}^{Capital}$ — Capital cost for electrolyzer per MW.
$c_{Comp}^{Capital}$ — Capital cost for hydrogen compressor per MW.
$c_{FC}^{Capital}$ — Capital cost for fuel cell per MW.
$c_{Cav}^{Capital}$ — Capital cost for salt cavern per kg.
$c_{WT}^{O\&M}$ — Operation and maintenance cost for wind turbine per year.
$c_{BESS}^{O\&M}$ — Operation and maintenance cost for BESS per year.
$c_{El}^{O\&M}$ — Operation and maintenance cost for electrolyzer per year.
$c_{FC}^{O\&M}$ — Operation and maintenance cost for fuel cell per year.
$c_{Cav}^{O\&M}$ — Operation and maintenance cost for cavern per year.
$P_{BESS}^{max}$ — Maximum charge/discharge power of BESS.
$P_{BESS}^{min}$ — Minimum charge/discharge power of BESS.
$P_t^{Load}$ — Load demand at time period $t$.
$P_{Max}^{Load}$ — Peak value of load demand.
$P_t^{WT}$ — Available wind power at time period t.
$P_{Rated}^{Rig}$ — Rated power of the offshore platform.
$T^E$ — Expected lifetime (years) of the planning system.
$E_{BESS}^{Initial}$ — Initial energy capacity of BESS.
$E_{Cav}^{Initial}$ — Initial energy capacity of hydrogen cavern.
$\eta_{BESS}^{Disc}$ — Discharge efficiency of BESS.
$\eta_{BESS}^{Char}$ — Charge efficiency of BESS.
$\eta_{FC}$ — Efficiency of fuel cell.
$\eta_{El}$ — Efficiency of electrolyzer.
$\varepsilon_h$ — Amount of energy per kilogram hydrogen.

*Variables*
$V_{WT}$ — Amount of wind turbine unit.
$V_{BESS}$ — Energy capacity of BESS (MWh).
$V_{El}$ — Power of electrolyzer (MW).
$V_{FC}$ — Power of fuel cell (MW).
$V_{Cav}$ — Size of hydrogen salt cavern.
$T^R$ — Resilience duration in hours.
$P_t^{Disc}$ — Discharging power of BESS $s$ in time period $t$.
$P_t^{Char}$ — Charging power of BESS $s$ in time period $t$.
$P_t^{El}$ — Power of electrolyzer in time period $t$.
$P_t^{FC}$ — Power of fuel cell in time period $t$.
$P_t^{Curt}$ — Renewable power curtailment in time period $t$.
$U_t^{Disc}$ — Discharging status of BESS determined at time period $t$. It is 1 if discharging status; otherwise 0.
$U_t^{Char}$ — Charging status of BESS determined at time period $t$. It is 1 if charging status; otherwise 0.
$E_{BESS}^t$ — Energy capacity level of BESS at time period $t$.
$E_{Cav}^t$ — Energy capacity level of cavern at time period $t$.

## I. Introduction

The Gulf of Mexico (GOM) is an important region for energy resource production. Roughly 15% of total U.S. crude oil production and 5% of total U.S. dry natural gas production come from the GOM region [1]. Gas and oil made up 55% of the world's CO2 emissions from fuel in 2019, and a significant proportion came from offshore oil and natural gas (O&G) platforms [2]. Offshore O&G rigs consume 16 terawatt-hours (TWh) energy to power their operations for a year [3].

The potential of US offshore wind power is vast and estimated to be more than 2,000 gigawatts (GW) [4]. Offshore wind turbines are typically much larger than onshore wind turbines. Offshore winds are stronger and more stable than onshore winds, which leads to higher capacity factors for offshore wind farms [5]-[8]. The US has set a national offshore wind goal of 30 GW by 2030 [9]. It is also worth noting that floating wind farms that can be placed in deep and ultra-deepwater regions are developing very fast.

Previous studies have developed some models for the wind power integrated offshore platforms. It has been proved that the integration of wind power and on site synchronous generators can keep the stability of system's voltage and frequency in a desired level in [10]-[12]. However, the wind power accounts for a small proportion of total power supply and the traditional generators are still dominating and producing large amount of CO2 emissions. The offshore platform that powered by the offshore wind farm and onshore grid with HVDC cable is proved as a reliable model in [13]. Therefore, the offshore platform increases the demand for onshore grid. Also, the model proposed in [13] is not self-sufficient even through there is no CO2 emissions. An independent offshore platform micrgorid model is developed and simulated in [14], but traditional generators are still considered in the model, which may not help achieve the reduction of CO2 emissions in the future. For the aforementioned wind power integrated offshore platforms, they either contain traditional generators as the main source and integrated with a small portion of wind power or are lack of energy storage systems to be self-sufficient and mitigate the fluctuation caused by the wind power. Also, none of them is powered by 100% renewable energy.

To bridge the aforementioned gaps, additional energy storage system is needed such as battery energy storage system (BESS) and hydrogen storage system (HESS). They are both very flexible and can mitigate the variation and intermittency of offshore renewable generation and improve the overall system reliability and resilience.

The deployment of utility-scale BESS has been increasing substantially in recent years. There was 1.5 GW operating battery power capacity in the US by 2020. Considering the planned battery storage projects, that number will rise to 16 GW by 2024. Batteries can achieve very high roundtrip energy conversion efficiencies of 90% or higher. Lithium-ion battery is the most widely used battery technology and it accounts for over 80% of U.S. large-scale battery storage due to its high energy density and low self-discharge rate [15].

HESS is also gaining popularity in recent years. One major driving force is the fast deployment of variable renewable generation while current power grids are unable to completely use up all those excess free and clean power. An effective solution selected by many nations to resolve this issue is to develop electrolyzers (El) that consume the excess power to produce hydrogen that can be stored in a high pressure tank with no self-discharge issue. Then, the hydrogen can be converted back into electricity either locally with co-located fuel cells (FC) or remotely with standalone FC where hydrogen is delivered to through pipelines. There are already many planned projects. For instance, the European Union has released their hydrogen strategy for a climate-neutral Europe that targets 40 GW of electrolyzers installed by 2030 [16].

This paper proposes a novel offshore hybrid renewable energy sources (OHRES) system with the integration of wind turbines (WTs), BESS and HESS to enable the offshore platform powered by 100% clean renewable energy with zero CO2 emissions. The HESS and BESS can effectivity mitigate the fluctuation caused by intermittent and stochastic wind power. Moreover, the resilience OHRES models proposed in this paper can ensure continuous power to the offshore demand at least for a pre-specified time interval in the worst scenario when wind power is completely cut off. In other words, the BESS and HESS are able to supply the load without the wind turbine and last for a certain time period. Therefore, the reliability and resilience of proposed OHRES system is further enhanced while CO2 emissions can be significantly reduced.

The remainder of the paper is organized as follows. The mathematical formulation for a basic OHRES model is presented in Section II. Section III describes the proposed resilience models in detail. Case studies and result discussions are presented in Section IV. Section V concludes the paper.

## II. OFFSHORE HYBRID RENEWABLE ENERGY SOURCES SYSTEM

The basic OHRES system consists of (1)-(18) as described below, which ensures power supply for the offshore platforms. The resilience is not considered in the basic model. The objective of the basic OHRES model is to minimize the lifetime total cost to maintain the power of the offshore platforms, which includes the cost of three subsystems: wind turbine, battery energy storage system, and hydrogen energy storage system; the objective function is presented in (1).

$$F(cost) = f(WT) + f(BESS) + f(HESS) \quad (1)$$

The total cost of HESS is shown in (2) which includes the cost of electrolyzer, fuel cell, compressor, and salt cavern. The hydrogen pipeline cost is not considered in this model since the OHRES system is local for an offshore platform.

$$f(HESS) = f(El) + f(FC) + f(Comp) + f(Cav) \quad (2)$$

Each subsystem's total cost is modeled with the capital cost, operation and maintenance cost and expected lifetime are shown in (3)-(8). Note that the power of the compressor is equal to the electrolyzer respectively.

$$f(WT) = V_{WT}(c_{WT}^{Capital} + c_{WT}^{O\&M} T^E) \quad (3)$$
$$f(BESS) = V_{BESS}(c_{BESS}^{Capital} + c_{BESS}^{O\&M} T^E) \quad (4)$$
$$f(El) = V_{El}(c_{El}^{Capital} + c_{El}^{O\&M} T^E) \quad (5)$$
$$f(FC) = V_{FC}(c_{FC}^{Capital} + c_{FC}^{O\&M} T^E) \quad (6)$$
$$f(Comp) = V_{El}(c_{Comp}^{Capital} + c_{Comp}^{O\&M} T^E) \quad (7)$$
$$f(Cav) = V_{Cav}(c_{Cav}^{Capital} + c_{Cav}^{O\&M} T^E) \quad (8)$$

The power balance equation involving renewable energy sources, BESS, HESS and load demand is shown below in (9):

$$P_t^{Disc} + V_{WT}P_t^{WT} + P_t^{FC} - P_t^{Load} - P_t^{Char} - P_t^{El} - P_t^{Curt} = 0, \forall t, \quad (9)$$

The BESS is model by (10)-(16) below. Constraint (10) calculates the stored energy of BESS for each time interval.

Equations (11)-(12) enforces the BESS initial energy level and the ending energy level. Constraint (13) forces the BESS energy level to be under the maximum limit of the BESS. Constraint (14) restricts the BESS in three modes: charging; discharging; or stay idle. Constraints (15)-(16) ensures the charging and discharging power are under the limits of BESS.

$$E_{BESS}^t - E_{BESS}^{t-1} = P_t^{Char}\eta_{BESS}^{Char} - P_t^{Disc}/\eta_{BESS}^{Disc}, \forall t, \quad (10)$$
$$E_{BESS}^{Initial} = 50\% V_{BESS}, \forall t, \quad (11)$$
$$E_{BESS}^{Initial} = E_{BESS}^{24}, \forall t, \quad (12)$$
$$0 \leq E_{BESS}^t \leq V_{BESS}, \forall t, \quad (13)$$
$$U_t^{Char} + U_t^{Disc} \leq 1, \forall t, \quad (14)$$
$$0 \leq P_t^{Disc} \leq U_t^{Disc} P_{Max}^{Disc}, \forall t, \quad (15)$$
$$0 \leq P_t^{Char} \leq U_t^{Char} P_{Max}^{Char}, \forall t, \quad (16)$$

The structure of HESS is modeled similar to BESS. However, the "charging" and "discharging" modes are achieved by electrolyzer and fuel cell respectively; and the energy is stored in the form of compressed hydrogen in the salt cavern. Constraint (17) calculates the stored energy of salt cavern at each time interval. Equations (18)-(19) enforces the initial energy level in the salt cavern and the ending energy level respectively. Constraint (20) forces the salt cavern energy level is under the limits of the HESS. Constraints (21)-(22) ensure the power outs of electrolyzer and fuel cell are under limits respectively. Equation (23) ensures the electrolyzer can refill the hydrogen in the salt cavern to full capacity in 24 hours.

$$E_{Cav}^t - E_{Cav}^{t-1} = P_t^{El}\eta_{El} - P_t^{FC}/\eta_{FC}, \forall t, \quad (17)$$
$$E_{Cav}^{Initial} = 50\% V_{Cav}, \forall t, \quad (18)$$
$$E_{Cav}^{Initial} = E_{Cav}^{24}, \forall t, \quad (19)$$
$$0 \leq E_{Cav}^t \leq V_{Cav}\varepsilon_h, \forall t, \quad (20)$$
$$0 \leq P_t^{El} \leq V_{El}, \forall t, \quad (21)$$
$$0 \leq P_t^{FC} \leq V_{FC}, \forall t, \quad (22)$$
$$V_{El} * 24 \geq V_{Cav}\varepsilon_h, \forall t, \quad (23)$$

### III. RESILIENCE MODELS

The basic OHRES system presented in the above section can ensure the system to maintain the power supply for the offshore platforms in a typical day. However, the stochastic and intermittent characteristic of offshore wind lead to different wind profiles in different days that may not maintain the power for offshore platforms with the basic model and may weaken the stability of the proposed OHRES system. Therefore, a resilience model is developed to ensure that, in the worst scenario when the wind power is completely cut off, the proposed system can continuously power the offshore platforms at least for a pre-specified time period. This time period is defined as the resilience duration $T^R$. Three resilience models are proposed to test and compare the performance of HESS, BESS and heterogeneous energy storage with BESS and HESS.

#### A. HESS Resilience Model

HESS resilience model is designed to maintain the demand only through HESS for $T^R$ during the worst scenario. Two constraints are formulated to archive the goal of suppling the load of the rig only through HESS. Constraint (24) ensures the output power of the fuel cell is greater than the peak load while (25) ensures the hydrogen energy stored in the salt cavern is enough to cover rated load during the resilience duration.

$$V_{FC} \geq P_{Max}^{Load} \quad (24)$$
$$V_{BESS}\eta_{BESS} \geq P_{Rated}^{Rig} T^R \quad (25)$$

#### B. BESS Resilience Model

In BESS resilience model, HESS will not be considered as a backup storage, instead, BESS will take over the responsibility of supplying the power to the offshore platforms when the wind power is not available. Similar to (24) and (25), (26) and (27) describe the power limit and energy limit respectively for BESS.

$$P_{Max}^{Disc} \geq P_{Max}^{Load} \quad (26)$$
$$V_{BESS}\eta_{BESS}^{Disc} \geq P_{Rated}^{Rig} T^R \quad (27)$$

#### C. Joint Resilience Model

In the joint resilience model, BESS and HESS are designed to power the offshore load together during the worst scenario. Constraints are formulated in (28) and (29) which are represented by the sum of the previous two models. Table I summarizes the proposed three resilience models.

$$P_{Max}^{Disc} + V_{FC} \geq P_{Max}^{Load} \quad (28)$$
$$V_{BESS}\eta_{BESS}^{Disc} + V_{Cav}\eta_{FC} \geq P_{Rated}^{Rig} T^R \quad (29)$$

TABLE I Formulations of Resilience Models

| Resilience Models | Constraints |
| --- | --- |
| A. HESS Resilience Model | (1)-(23), (24)-(25) |
| B. BESS Resilience Model | (1)-(23), (26)-(27) |
| C. Joint Resilience Model | (1)-(23), (28)-(29) |

### IV. CASE STUDIES

A typical 50 MW offshore platform is applied as a test bed to evaluate the proposed OHRES system. The traditional model that uses diesel generators to power the offshore platform is set as a benchmark model. The load profile of the test bed offshore platform is shown in Fig. 1. The load profile of the offshore platform does not fluctuate like the residential load since it operates 24 hours and 80% of the load is from mining and oil processing [17].

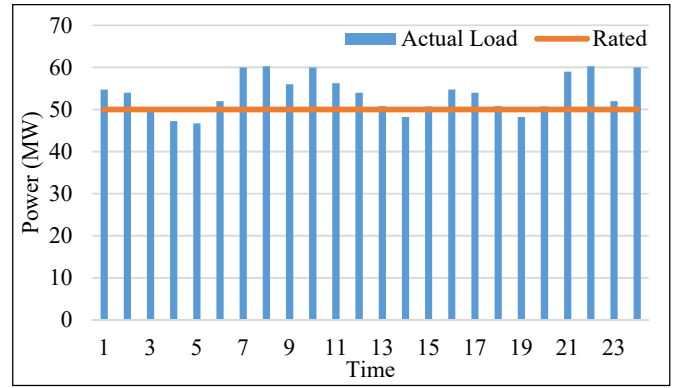

Figure 1. Load profile of offshore platform.

The offshore WTs rated at 3 MW each are modeled in this case studies. Different from the onshore WTs, the offshore WTs cost much higher on the construction part. The wind power is scaled by the data from [18] for a typical day at GOM. The wind profile in Fig. 2 represents the available wind power for each WT. Also, the initial tests are based on the pre-set parameters shown in Table II including the costs for WTs, BESS, electrolyzer, fuel cell, compressor and salt cavern. The charging and discharging efficiency for BESS are set to 90% while it is 70% for electrolyzer and 60% for fuel cell respectively.

The proposed OHRES basic model and the resilience models are all solved by the python package "Pyomo" [19] and "Gurobi" optimizer solver [20].

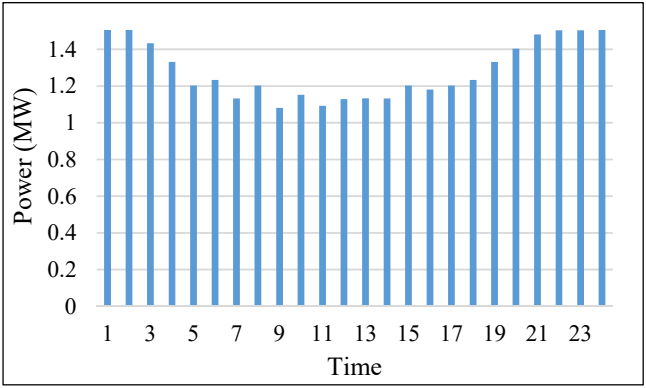

Figure 2. Wind power profile for a 3 MW unit.

Table II Cost parameters.

| System | WT (3MW) | BESS (MWh) | El (MW) | FC (MW) | Comp (MW) | Cavern (Ton) |
|---|---|---|---|---|---|---|
| Cost \ million $ | 20 | 0.35 | 1.2 | 1 | 0.04 | 0.035 |

Table III Results with $T^R$ of 6 hours

| Models | Basic | A | B | C |
|---|---|---|---|---|
| WT | 42 | 43 | 42 | 42 |
| BESS (MWh) | 77.39 | 34.19 | 315.79 | 74.18 |
| El (MW) | 0.16 | 29.76 | 0 | 22.77 |
| FC (MW) | 0.18 | 60.25 | 0 | 41.71 |
| Cavern (kg) | 33.64 | 15151.5 | 0 | 11592.3 |
| Capital ($ in Million) | 867.51 | 986.62 | 950.53 | 949.29 |
| Operation ($ in Million) | 51.78 | 73.80 | 99.33 | 72.79 |
| Total ($ in Million) | 919.28 | 1060.42 | 1049.86 | 1022.08 |
| Average ($/MWh) | 104.94 | 121.05 | 119.85 | 116.68 |

Table IV Results with $T^R$ of 12 hours

| Models | Basic | A | B | C |
|---|---|---|---|---|
| WT | 42 | 43 | 42 | 42 |
| BESS (MWh) | 77.39 | 34.19 | 631.58 | 74.18 |
| El (MW) | 0.16 | 59.52 | 0 | 52.53 |
| FC (MW) | 0.18 | 60.25 | 0 | 41.71 |
| Cavern (kg) | 33.64 | 30303.0 | 0 | 26743.8 |
| Capital ($ in Million) | 867.51 | 1041.03 | 1061.05 | 1003.7 |
| Operation ($ in Million) | 51.78 | 88.09 | 162.49 | 87.07 |
| Total ($ in Million) | 919.28 | 1129.12 | 1223.54 | 1090.77 |
| Average ($/MWh) | 104.94 | 128.89 | 139.67 | 124.52 |

Tables III-VI represent the results of the optimal 20-year planning cost for the OHRES basic and resilience models based on different length of $T^R$. The average cost is calculated based on the total electricity consumption for a typical 50MW offshore platform that would consume 8,760,000 MWh in 20 years. From the results, we can observe that with the increase of $T^R$, the average electricity cost increases for all models A, B and C. However, the costs for the basic model remain the same because the resilience is not considered there. Comparing with model A and model B, we can find that the costs for model A is higher than model B when $T^R$ equals to 6 hours. If the $T^R$ is greater than 6 hours, the costs for model A is lower than model B. This indicates that HESS has advantages when large amount of energy capacity is planned. For the joint system, model C has the best performance with least costs among all three resilience models. The OHRES results shown in Table VII is the results of model C from Table III. From the results, the capital cost accounts for a majority portion of the total cost in the proposed OHRES system, while the operation cost (mainly fuel cost) is the dominating cost for traditional systems. The $CO_2$ emissions can be reduced from 526,000 tons to zero if the traditional system is replaced by the proposed OHRES system on a 50 MW offshore platform.

Table V Results with $T^R$ of 18 hours

| Models | Basic | A | B | C |
|---|---|---|---|---|
| WT | 42 | 43 | 42 | 42 |
| BESS (MWh) | 77.39 | 34.19 | 947.37 | 74.18 |
| El (MW) | 0.16 | 89.29 | 0 | 82.29 |
| FC (MW) | 0.18 | 60.25 | 0 | 41.71 |
| Cavern (kg) | 33.64 | 45454.5 | 0 | 41895.3 |
| Capital ($ in Million) | 867.51 | 1095.43 | 1171.58 | 1058..1 |
| Operation ($ in Million) | 51.78 | 102.38 | 225.65 | 101.36 |
| Total ($ in Million) | 919.28 | 1197.81 | 1397.23 | 1159.46 |
| Average ($/MWh) | 104.94 | 136.74 | 159.50 | 132.36 |

Table VI Results with $T^R$ of 24 hours

| Models | Basic | A | B | C |
|---|---|---|---|---|
| WT | 42 | 43 | 42 | 42 |
| BESS (MWh) | 77.39 | 34.19 | 1263.16 | 74.18 |
| El (MW) | 0.16 | 119.05 | 0 | 112.06 |
| FC (MW) | 0.18 | 60.25 | 0 | 41.71 |
| Cavern (kg) | 33.64 | 60606.1 | 0 | 57046.8 |
| Capital ($ in Million) | 867.51 | 1149.84 | 1282.11 | 1112.51 |
| Operation ($ in Million) | 51.78 | 116.66 | 288.81 | 115.64 |
| Total ($ in Million) | 919.28 | 1266.50 | 1570.91 | 1228.15 |
| Average ($/MWh) | 104.94 | 144.58 | 179.33 | 140.20 |

Table VII Comparisons between OHRES and traditional system

| Model | Capital (Million) | Operation (Million) | Total (Million) | Average ($/MWh) | Emission (tons) |
|---|---|---|---|---|---|
| Traditional | 12.5 | 788.4 | 800.9 | 91.42 | 526,500 |
| OHRES | 949.29 | 72.79 | 1022.08 | 116.68 | 0 |

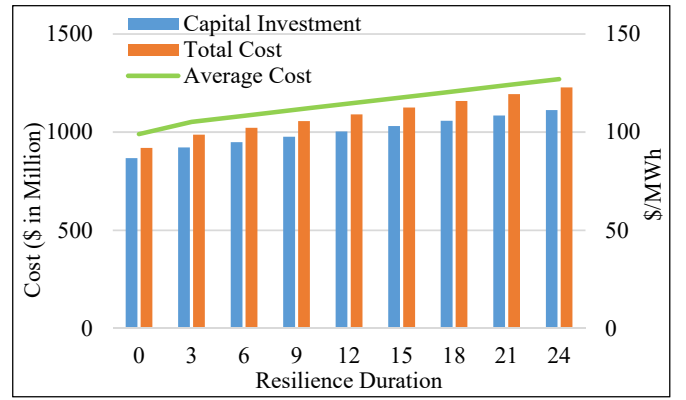

Figure 3. Sensitivity test of $T^R$ for Joint resilience model.

Fig. 3 present the results of different $T^R$ for the joint resilience model with the parameters shown in Table II. The difference between the capital investment cost and the total cost is the operation cost for the 20 years planning horizon. From Fig. 3, we can observe that all the costs increase with the increase of $T^R$. The results match the previous Tables because a longer period of $T^R$ will increase the energy storage capacity. In other words, the capital cost of the BESS and/or HESS increases at the same time. A longer period of $T^R$ can significantly increase the reliability and resilience of the proposed OHRES system, which increases the total cost. A tradeoff between resilience and economics must be made.

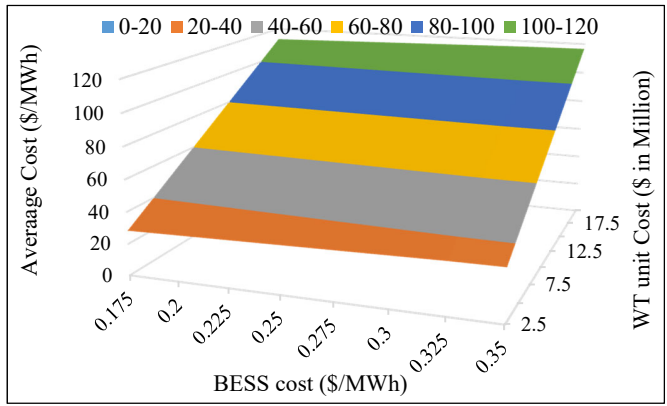

Figure 4. Average electricity price with different cost of BESS and WT.

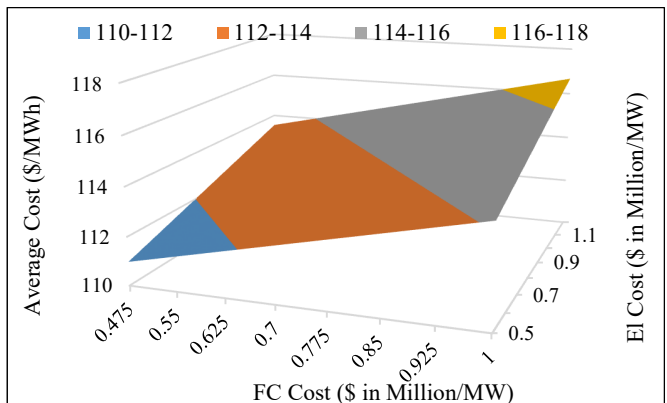

Figure 5. Average electricity price with different cost of El and FC.

Figures 4-5 present the sensitivity tests of the average electricity cost based on different costs of the renewable units in BESS or HESS. The unit cost of El and FC in Fig. 4 are set to constant per Table II. Similarly, the costs of WT and BESS in Fig. 5 are set to constant per the same table. The average electricity cost is higher than the traditional system under the current costs of renewable units. However, most of the costs of renewable units are projected to decrease in the next 10 years. From the results, we can conclude that with the decreasing cost of renewable units, the average electricity cost for the OHRES system will be lower than the traditional system eventually.

Compare to BESS, HESS does not degrade significantly. Therefore, the BESS degradation cost is considered in $c_{BESS}^{O\&M}$ as 5% of the BESS capital cost per year. This might be the reason that the HESS is more competitive than BESS when large capacity of energy storage is planned. However, the linear degradation cost that applied in this paper may not accurately represent the BESS degradation. [21]-[22] present a neural network based battery degradation model that can accurately predict the degradation of BESS which can be leveraged to improve the OHRES planning model in future.

## V. CONCLUSION

A novel renewable energy sources system is proposed to replace the traditional electricity generation to power the offshore platforms by 100% clean renewable energy with zero $CO_2$ emissions. Three resilience models are designed to enhance the reliability of the proposed OHRES system by imposing extra power capability and energy storage of BESS and/or HESS. The total cost of OHRES system increases with a higher level of resilience. Based on the current unit costs, the average electricity cost of the proposed OHRES system is higher than the traditional onsite diesel generators. However, the $CO_2$ emissions from offshore platforms can be eliminated each year with the proposed OHRES system. Also, the statistical economic results of the proposed system give an insight regarding when the proposed OHRES system will be more economical than the traditional diesel generators with the decreasing cost of renewable units. To summarize, the proposed OHRES system is feasible to replace the traditional system while maintaining the system reliability, and its zero emission benefits largely contribute to global decarbonization.